\documentclass[final,leqno,onefignum,onetabnum]{siamonline0516}%
\usepackage{amsmath}
\usepackage{amsfonts}
\usepackage{amssymb}
\usepackage{graphicx}
\usepackage{afterpage}
\usepackage{float}
\usepackage{caption}
\usepackage{subcaption}
\usepackage{hyperref}
\begin{document}

\title{Decision Dynamics in Groups with Interacting Members\thanks{Submitted on August 23, 2016}}
\author{Reginald J Caginalp\thanks{Department of Mathematics, University of Pittsburgh, Pittsburgh, PA, 15260, USA.} \and Brent Doiron\footnotemark[2] \thanks{Center for the Neural Basis of Cognition, Pittsburgh, PA, 15213, USA. }}
\maketitle

\begin{abstract}
Group decisions involve the combination of evidence accumulation by individual members and direct member-to-member interactions.  We consider a simplified framework of two deciders, each undergoing a two alternative forced choice task, with the choices of early deciding members biasing members who have yet to choose.  We model decision dynamics as a drift-diffusion process and present analysis of the associated Fokker-Planck equation for the group.  We show that the probability of coordinated group decisions (both members make the same decision) is maximized by setting the decision threshold of one member to a lower value than its neighbor's.  This result is akin to a speed-accuracy tradeoff, where the penalty of lowering the decision threshold is choice inaccuracy while the benefit is that earlier decisions have a higher probability of influencing the other member.  We numerically extend these results to large group decisions, where it is shown that by choosing the appropriate parameters, a small but vocal component of the population can have a large amount of influence on the total system.

\end{abstract}

\slugger{siads}{xxxx}{xx}{x}{x--x}



\section{Introduction}

There is a long history of study in how evidence is integrated and ultimately drives decisions \cite{edwards1954,Saaty1990,bogacz2006}.  An often used framework is the {\em two-alternative forced choice task} (TAFC), where decisions are constrained to be between two alternatives with evidence steadily accumulated over time.  While the TAFC framework is admittedly oversimplified it has provided a wealth of data by which to compare and contrast various models of decision processes \cite{bogacz2006}.   Many models treat TAFC decision dynamics as a drift-diffusion stochastic process \cite{Vickers1970,ratcliff1978,usher2001,bogacz2006,ratcliff2008}, where the drift term models the steady accumulation of evidence and the diffusion term models variability in decision making. Drift-diffusion models capture observed decision behavior at both single neuron \cite{gold2007} and psychophysical levels \cite{ratcliff2016}, and has been shown to perform optimal decision making with appropriate assumptions \cite{bogacz2006}.  

Decisions are not always made by individuals in isolation, yet by members in a group where all individuals are actively engaged in the decision process \cite{yaniv2004,couzin2009}. There are mixed reports about the benefits (or lack of) of deciding within a group, with examples where group decisions are more accurate \cite{ward2011} and others where a systematic group bias is detected \cite{davis1992}. Nevertheless, group decision making theory has been applied to economics, political science, and animal behavior \cite{banerjee1992,black1948,conradt2003,ben2000,goyal2012,couzin2009,sorkin2001}.  Combining the drift-diffusion dynamics of a population of deciders performing a TAFC task to form a {\it group decision} is a natural extension, and several groups have made important advances in this area (see \cite{kimura2012} for a review).   However, these modeling studies rarely consider interactions between deciders during evidence accumulation (but see \cite{srivastava2014}). 

In this paper we consider a group of deciders each engaged in a TAFC task with each member modeled as a drift-diffusion processes.  When a member of the group makes a decision it communicates its choice to all other members of the group, with the hope to influence those who have yet to decide. In general, introducing coupling between drift-diffusion processes creates significant challenges in any analysis. To make our model tractable we consider an instantaneous interaction: when a certain member decides, it `kicks' all the other members towards the decision it made. Following the interaction, the decision makers continue their drift-diffusion processes independent of one another. Therefore, other than at a finite set of interaction times, the stochastic processes are independent of each other. This type of interaction permits a calculation of the probabilities of group decisions through extensions of the single decider framework.

We begin with the simple case of two deciders. One of the deciders is biased towards the + choice (without loss of generality). By varying the amount of evidence that is required for this decider to make a decision (decision threshold), we aim to maximize the probability that both deciders choose the + decision (++ decision). In the case where there is no interaction between the deciders, the optimal solution for the + decider is to require a very large amount of evidence to make a decision. In this way, the random noise from the diffusion term is irrelevant and the + decider always makes the + decision.  Thus, the probability of a joint ++ decision rests solely on the other decider (which the $+$ decider has no influence over). We contrast this to the case with decider interaction.  We find that for large enough decider coupling a finite decision threshold maximizes the probability of a ++ decision. A compromise occurs between the + decider sometimes choosing the $-$ decision in error, but the + decider having the possibility of influence over the other decider.   We conclude by showing that the intuition gained for the two decider case carries over to a large $N$-decider population.

\section{Results}

\subsection{Stochastic dynamics in pairs of deciders}

\begin{figure}[h!]
\centering
  \includegraphics[width=0.9\textwidth]{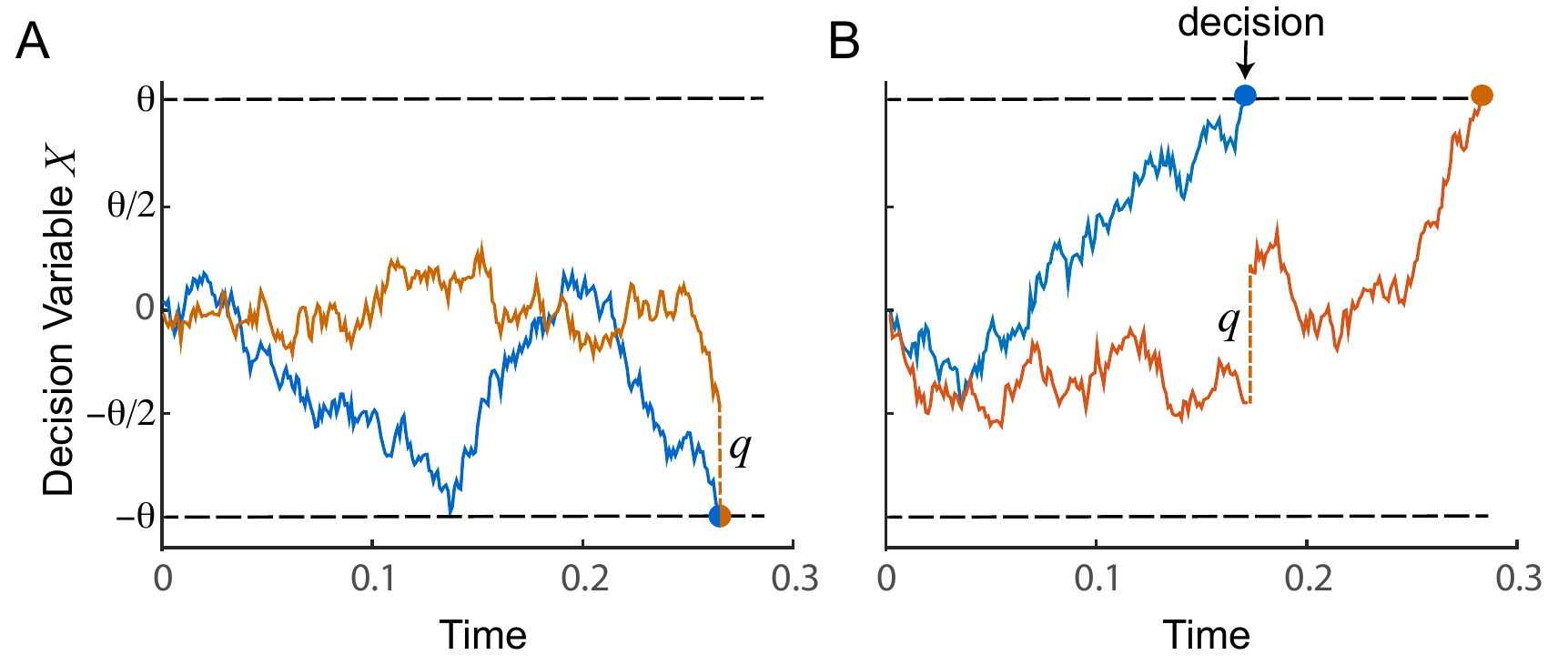}
    \caption{Modeling group decisions with drift-diffusion dynamics. \textbf{A.} Instantaneous kick: A decider crosses through the $-$ threshold, and immediately kicks its neighbor across the same threshold. \textbf{B.} Kick and diffusion: A decider  crosses through the $+$ threshold and kicks its neighbor up by an amount $q=\theta/2$ after which the decider then diffuses across the $+$ threshold.}
     \label{Fig1}
\end{figure}

We begin by considering a system of two deciders, each of which is trying to decide between a + choice and a -- choice (Figure \ref{Fig1}).  The group decision dynamics obey the following pair of Langevin equations \cite{risken1984}: 
\begin{eqnarray}
dX_{1}&=&\mu_1 dt+\sqrt{2D}dW_{1}+G_2q\delta(t-t_1), \label{MC1} \\%
dX_{2}&=&\mu_2 dt+\sqrt{2D}dW_{2}+G_1q\delta(t-t_2). \label{MC2}
\end{eqnarray}
The processes $X_1(t)$ and $X_2(t)$ represent the amount of evidence collected by the decider 1 and 2 at time $t$, respectively.   For a given decider $i$ ($i = 1,2$), when the evidence $X_i(t)$ reaches a value $+\theta_i$ it decides on the + choice, but if the evidence reaches $-\theta_i$, it decides on the $-$ choice.  
We consider evidence accumulation to be stochastic and include the Brownian processes $W_1(t)$ and $W_2(t)$ ($W_1(t)$ and $W_2(t)$ are statistically independent), with diffusion coefficient $D>0$.  Finally, we set $X_1(0)=X_2(0)=0$ so that neither decider is initially biased towards the $+$ or $-$ decision.

The random decision time for decider $i$ is denoted by $t_i$ and $G_i=1$ if $i$ chooses the + choice, while $G_i=-1$ if $i$ chooses the $-$ choice. The interaction between the deciders is modeled by the final terms in Eqs. \eqref{MC1} and \eqref{MC2}.  When decider $i$ chooses the $\pm$ choice at time $t_i$, it provides an instantaneous evidence kick of intensity $G_iq$ at time $t_i$ to $X_j$ ($j \ne i$). In other words, upon a decision a decider will attempt to influence its neighbor to choose the same choice it did. We remark that the coupling term is only relevant if the neighbor decider has yet to decide.  Assuming decider $i$ decides before decider $j$, then the interaction can separated into two cases: either decider $i$ kicks $j$ across $\pm \theta_j$ instantly (meaning $t_i=t_j$; see Figure \ref{Fig1}A), or it kicks decider $j$ and it eventually drifts across one of the boundaries at a later time ($t_j > t_i$;  see Figure \ref{Fig1}B).  An alternative model of decision coupling would be for $\mu_j \to \mu_j +q$ at time $t_i$.  This model would imply that a neighbor's decision is a continual source of evidence to the other decider, which must be accumulated over time to have influence.  In this study we confine ourselves to the former model where decisions are communicated in an instant. 

We are interested in the group decision ($G_1,G_2$). One approach for obtaining the probabilities of ($G_1,G_2$) is to estimate them from many Monte-Carlo realizations of Eqs. \eqref{MC1} and \eqref{MC2}. Another is to solve an associated Fokker-Planck equation to obtain analytic estimates for  $G_1$ and $G_2$. We follow both of these approaches and find that they agree very closely. 

\subsection{Calculating group decisions with interaction}
\label{sec:Inter_form}
The group decision dynamics within our model is a two dimensional problem governed by the concentration $c(x_1,x_2,t)$.  With proper normalization (see below) the concentration is the probability density at time $t$ for the evidences $X_1$ and $X_2$ over $x_1 \in (-\theta_1,\theta_1)$ and $x_2\in (-\theta_2,\theta_2)$, respectively. Since $X_1$ and $X_2$ are independent before the first interaction at time $t_i$, the evolution of the system decouples for $t < t_i$ and we get that $c(x_1,x_2,t)=c_1(x_1,t)c_2(x_2,t)$. The stochastic dynamics of either Eqs. \eqref{MC1} and \eqref{MC2} obey the associated Fokker-Planck equation \cite{risken1984} for $c_i(x,t)$:
\begin{equation}
\frac{\partial c_i}{\partial t}=-\mu_i\frac{\partial c_i}{\partial x}+D\frac
{\partial^{2}c_i}{\partial x^{2}}, \hspace{1.0cm} i=1,2. \label{BM2}%
\end{equation}
A decision occurs when $X_i$ reaches one the boundaries $\pm\theta_i$; this amounts  to supplementing Eq. \eqref{BM2} with the absorbing
boundary conditions:
\[
c_i\left(  \theta_i,t\right)  =c_i\left(  -\theta_i,t\right)  =0,
\]
for all $t>0.$ Furthermore, there is no evidence accumulation for $t < 0$ so that the
concentration at time $t=0$ obeys:%
\[
c_i\left(  x,0\right)  =\delta\left(  x\right)  .
\]

In general, under these conditions the drift-diffusion equation admits the Fourier series solution: %
\begin{equation} \label{eq:Fourier}
c_i\left(  x,t\right)  =\sum_{m=1}^{\infty}\frac{e^{_{\frac{\mu_i x}{2D}}}%
}{2\theta_i}\left(  -1\right)  ^{m+1}e^{-k_{2m-1}t}\sin\left(  w_{2m-1}\left(
x+\theta_i\right)  \right)
\end{equation}
 where
\begin{align*}
w_{2m-1}  &  \equiv\frac{\left(  2m-1\right)  \pi}{2\theta_i},\\
k_{2m-1}  &  \equiv\frac{\mu_i^{2}}{4D}+Dw_{2m-1}^{2}.
\end{align*}

\begin{figure}
\centering
  \includegraphics[width=0.9\textwidth]{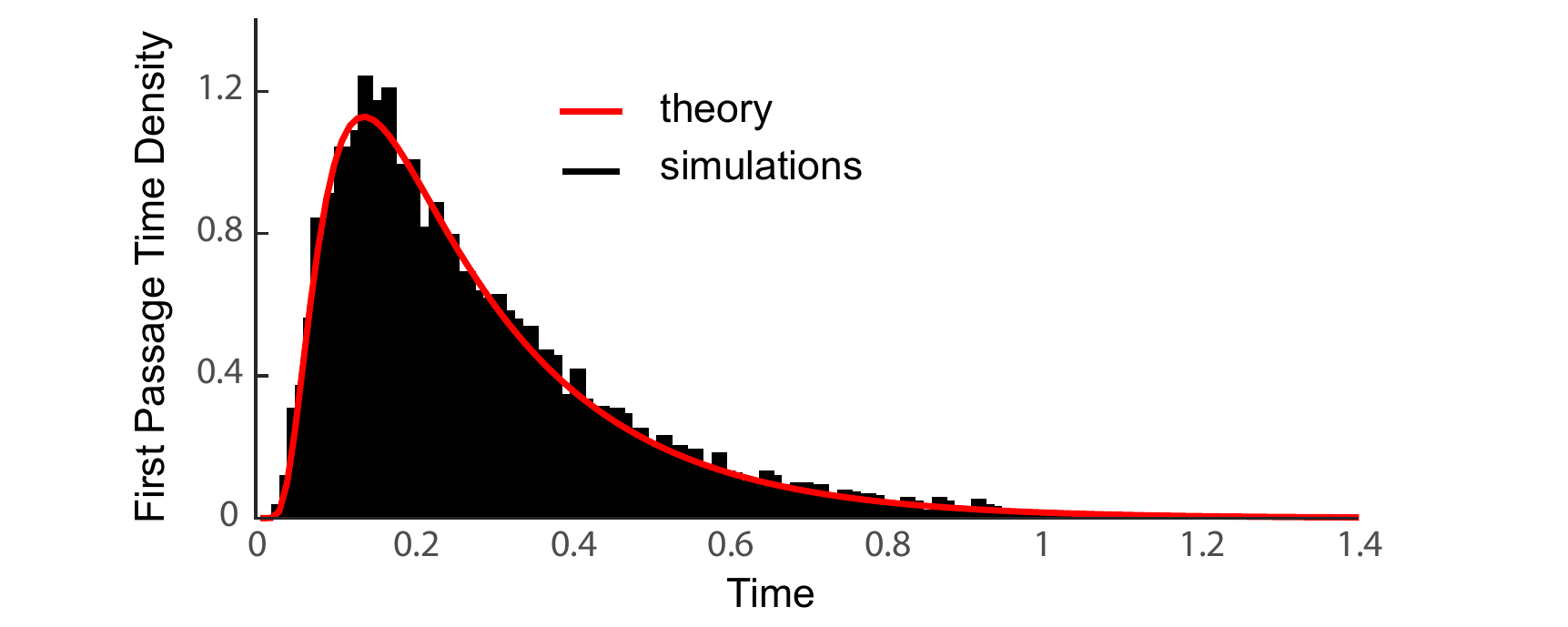}

    \caption{Simulated and theoretical FPT densities for decider 1 making the + decision first. We take $D=1$, $\mu_1=-\mu_2=0.75$, and $\theta_1=\theta_2=1$.  For the theoretical FPT density we truncated the Fourier series solution for $c(x,t)$ at mode $m=100$.}
     \label{Fig2}
\end{figure}

In what follows we wish to calculate the probability that both deciders cross the $+$ threshold (without loss of generality).  This requires incorporating the interaction that happens at the decision time of the first decider.  
The conditioned first passage time (FPT) density of
decider $i$ is denoted $f_i^{\pm }(t)$, and it describes the probability
that the decider will make the $\pm$ decision at time $t$. Note
that $\int_{0}^{\infty} f_i^{\pm }(t) dt$ is the total probability that the
decider makes the $\pm$ choice, and we have that $\int_{0}^{\infty}
[f^{+ }_i(t)+f^{- }_i(t)] dt = 1$ (i.e there is always a decision). The total FPT density of decider $i$ is then:
\[
f_i(t) \equiv f_i^{+}(t)+f_i^{- }(t).
\]
The FPT densities can be computed from the flux of concentration passing the threshold at time $t$:
\[
f^{\pm }_i\left(  t\right)    =\mp \left. D\frac{\partial c}{\partial x}  \right \vert
_{x=\pm\theta_i}.
\]
If we condition on decider $i$ deciding before $j$ ($t_i < t_j$) then the FPT density
of decider $i$ escaping through the $\pm\theta$
threshold at time $t_i$ is simply:
\begin{equation} \label{eq:cFPT}
f_{i}^{\pm }(t_{i}\vert t_i < t_j) = f_{i}^{\pm}(t_{i})\int_{t_{i}}^{\infty}  f_{j}(t_{j})
dt_{j}.
\end{equation}
As expected the conditioned $f_{i}^{\pm \theta}(t_{i}\vert t_i < t_j)$ is an asymmetric, single mode function (Figure \ref{Fig2}).  Using the truncated Fourier series solution in Eq. \eqref{eq:cFPT} gives an excellent agreement for $f_{i}^{\pm}(t_{i}\vert t_i < t_j)$ estimated from direct simulations of Eqs. \eqref{MC1} and \eqref{MC2} (Figure \ref{Fig2}).

Immediately before decider $j$ is kicked at time $t_{i}$, labelled $t_{i}^{\rightarrow}$ \footnote{To avoid cumbersome notation we denote the left (right) limit $ t \to t_i$ as $t_i^{\rightarrow}$ ($t_i^{\leftarrow}$).}, the concentration
$c_{j}(x,t_i^\rightarrow)$ of decider $j$ is given by:
\[
c_{j}(x,t_i^{\rightarrow})=\sum_{m=1}^{\infty}\frac{e^{_{\frac{\mu_{j}x}{2D}}}}%
{2\theta_{j}}\left(  -1\right)  ^{m+1}e^{-k_{2m-1}t_{i}}\sin\left(
w_{2m-1}\left(  x+\theta_{j}\right)  \right)  .
\]
For decider $j$ to escape at the $\pm$ threshold, there are two cases to consider: 1) $j$ is
 kicked across the $\pm$ threshold instantly ($t_j=t_i$; Figure \ref{Fig1}A), or 2) $j$ is kicked and
then diffuses across the $\pm$ threshold at a later time ($t_j>t_i$, Figure \ref{Fig1}B).
The conditional probability that $j$ crosses the $\pm$ gate can be thus decomposed as: 
\begin{multline*}
P(G_j = \pm 1|i\text{ crossing }\pm\text{ gate at
time }t_{i}) \\
= P(j\text{ inst. crossing }\pm\text{ gate}|i\text{ crossing }\pm\text{ gate at
time }t_{i}) \\ + P(j\text{ diff. across }\pm\text{ gate}|i\text{ crossing }\pm\text{ gate at
time }t_{i}).
\end{multline*}

If $X_j$ is between the lower value $L_{j}^{+}\equiv\theta_{j}-q$ and the upper value
$U_{j}^{+}\equiv\theta_{j}$, then the decider will be kicked instantaneously across the
$+\theta_{j}$ gate (assuming decider $i$ made the + choice). Similarly, if the decider $j$ is between $L_{j}%
^{-}\equiv-\theta_{j}$ and $U_{j}^{-}\equiv-\theta_{j}+q$, then the decider will be kicked
instantaneously across the $-\theta_{j}$ gate (assuming decider
$i$ made the - choice). The probability of
instantaneous crossing conditioned on $i$ crossing the $\pm$ gate at time
$t_{i}$ is then the probability that the evidence $X_j$ will be in the range
$(L_{j}^{\pm},U_{j}^{\pm})$. That is,
\[
P(j\text{ inst. crossing }\pm\text{ gate}|i\text{ crossing }\pm\text{ gate at
time }t_{i})=\frac{\int_{L_{j}^{\pm}}^{U_{j}^{\pm}}c_{j}(x,t_i^{\rightarrow})dx}%
{\int_{-\theta_{j}}^{\theta_{j}}c_{j}(x,t_i^{\rightarrow})dx}.
\]
We can make this expression simpler by defining the density $\rho(x,t)$ as the normalized concentration:
\[
\rho_i(x,t)\equiv\frac{c_i(x,t)}{\int_{-\theta_i}^{\theta_i}c_i(x,t)dx}.
\]
This means that $\int_{-\theta_i}^{\theta_i}\rho_i(x,t)dx=1$ for all $t>0$. Then the
above equation becomes
\begin{equation}
P(j\text{ inst. crossing }\pm\text{ gate}|i\text{ crossing }\pm\text{ gate at
time }t_{i})=\int_{L_{j}^{\pm}}^{U_{j}^{\pm}}\rho_{j}(x,t_i^{\rightarrow})dx. \label{eq:inst}
\end{equation}

We next must treat the case when decider $j$ is kicked and then diffuses across the threshold separately.   At time $t_{i}$ decider $j$ is kicked with magnitude $q$ and we assume that $X_j(t_i^{\rightarrow}) \pm q \in (-\theta_j,\theta_j)$.  Immediately after the kick the concentration $c_{j}(x,t_{i}^+)$ of decider $j$ is shifted by magnitude
$q$ in the $\pm x$ direction:
\[
c_{j}(x,t_i^{\leftarrow}) = c_{j}(x \mp q,t_i^{\rightarrow}).
\]
For times $t>t_i$ we again have independent diffusion and the evidence accumulation of decider $j$ can be obtained from the one dimensional diffusion process (Eq. \eqref{BM2}), now with an initial density of $c_{j}(x,t_i^{\leftarrow})$ (as opposed to $c_j(x,0)=\delta(x)$).  However, in this case we are only interested in the decision $G_j$ and not the decision time $t_j$.  

For simple random walk dynamics it is well known that if decider $j$ has evidence $x$ the probability that it will cross through
the $\pm$ gate is denoted by $\epsilon^{\pm}_j$ is given by \cite{redner2001,wald1947}: 
\begin{eqnarray*}
\epsilon^{+}_j(x) &=& \exp\left( \frac{\mu_j (\theta_j-x)}{2D}\right) \frac
{\sinh[\mu_j(x+\theta_j)/(2D)]}{\sinh[2\mu_j \theta_j/(2D)]}, \\
\epsilon^{-}_j(x) &=& 1-\epsilon^{+}_j(x).
\end{eqnarray*}

If $i$ escapes through the positive gate then decider $j$ is not kicked across
instantly if $X_j$ is between $\Lambda_{j}^{+} \equiv-\theta_{j}$ and
$\Omega_{j}^{+} \equiv\theta_{j}-q$. Equivalently, if $i$ escapes through the negative
gate, decider $j$ is not kicked across instantly if it is between
$\Lambda_{j}^{-} \equiv-\theta_{j}+q$ and $\Omega_{j}^{-} \equiv
\theta_{j}$. From this we have the probability that $j$ is not kicked across instantaneously being $\int_{\Lambda_{j}^{\pm}}^{\Omega_{j}^{\pm}} \rho_{j}(x,t_i^{\rightarrow})
dx$. Once decider $j$ has been kicked its probability density is given by
$\rho_{j}(x,t_i^{\leftarrow})$. Thus, the total conditional probability that decider $j$ diffuses through the
$\pm$ gate is:
\begin{multline}
P(j \text{ diff. across } \pm\text{ gate} \vert i \text{ crossing } \pm\text{
gate at time } t_{i}) \\ =\int_{\Lambda_{j}^{\pm}}^{\Omega_{j}^{\pm}} \rho
_{j}(x,t_i^{\rightarrow}) dx \int_{-\theta_{j}}^{\theta_{j}} \rho_{j}(x,t_i^{\leftarrow})
\epsilon^{\pm }_{j}(x) dx \label{eq:diff}.
\end{multline}

Finally, the probability of $j$ crossing the $\pm$ gate (by whatever means) conditioned
on $i$ crossing the same gate at $t_{i}$ is from Eqs. \eqref{eq:inst} and \eqref{eq:diff}:
\begin{multline}
P(G_j = \pm 1 \vert i \text{ crossing } \pm\text{ gate
at time } t_{i}) \\ = \int_{L_{j}^{\pm}}^{U_{j}^{\pm}} \rho_{j}(x,t_i^{\rightarrow})dx +
\int_{\Lambda_{j}^{\pm}}^{\Omega_{j}^{\pm}} \rho_{j}(x,t_i^{\rightarrow}) dx
\int_{-\theta_{j}}^{\theta_{j}} \rho_{j}(x,t_i^{\leftarrow}) \epsilon^{\pm }_{j}(x) dx. 
\end{multline}
Putting this all together, we find that the
probability that both deciders cross the $\pm$ gate (with $i$ crossing before $j$) is
given by:
\begin{multline}
P(G_i=G_j =\pm 1 \vert t_i < t_j ) \\ = \int_{0}^{\infty}  f_{i}^{\pm }(t_{i}\vert t_i < t_j) \Bigg[  \int_{L_{j}^{\pm}}%
^{U_{j}^{\pm}} \rho_{j}(x,t_i^{\rightarrow})dx \\ + \int_{\Lambda_{j}^{\pm}}^{\Omega
_{ij}^{\pm}} \rho_{j}(x,t_i^{\rightarrow}) dx \int_{-\theta_{j}}^{\theta_{j}} \rho
_{j}(x,t_i^{\leftarrow}) \epsilon^{\pm }_{j}(x) dx \Bigg]  dt_{i}. \label{eq:xx}
\end{multline}

Without loss of generality we now calculate the probability that both decide on the + choice. In this case, we have that Eq. \eqref{eq:xx} is: 
\begin{multline*}
P(G_i=G_j = 1 \vert t_i < t_j ) \\ =\int_{0}^{\infty} f_{i}^{+}(t_{i}\vert t_i < t_j)  \Bigg[  \int_{\theta_{j}-q}^{\theta_{j}}\rho
_{j}(x,t_i^{\rightarrow})dx \\ +\int_{-\theta_{j}}^{-\theta_{j}+q}\rho_{j}%
(x,t_i^{\rightarrow})dx\int_{-\theta_{j}}^{\theta_{j}}\rho_{j}(x,t_i^{\leftarrow})\epsilon^{+
}_{j}(x)dx\Bigg]  dt_{i}.
\end{multline*}

Finally, the probability of a $+,+$ group decision is then given by
the sums of the probabilities for the cases that $t_1 < t_2$ and $t_2 < t_1$.  This yields:
\begin{multline}
P_{++}  \equiv \textrm{Prob}(G_1=1,G_2=1)  \\ = \sum_{\substack{i=1\\j\neq i}}^2\int_{0}^{\infty}f_{i}^{ + }(t_{i}\vert t_i < t_j) \Bigg[  \int_{\theta_{j}-q}^{\theta_{j}}%
\rho_{j}(x,t_i^{\rightarrow})dx \\ + \int_{-\theta_{j}}^{-\theta_{j}+q}\rho
_{j}(x,t_i^{\rightarrow})dx\int_{-\theta_{j}}^{\theta_{j}}\rho_{j}(x,t_i^{\leftarrow}%
)\epsilon^{+ }_{j}(x)dx\Bigg]  dt_{i}. \label{eq:final}
\end{multline}

Eq. \eqref{eq:final} naturally decomposes into two terms.  The first term represents the contribution whereby decider $j$ is kicked across the threshold instantly from the kick that it receives from decider $i$. The second term is the contribution when $j$ is kicked by $i$, and then $j$ diffuses, at a later time, across the threshold.

\subsection{Simulation and theory results}

To explore the joint decision dynamics of Eqs. \eqref{MC1}-\eqref{MC2} we begin by setting $D=1,$
$\mu_1=-\mu_2=0.75,$ $\theta_{2}=1$, and varying over $\theta_{1}$ and $q$. More to the point, we assume that the deciders have opposing drifts so that if $D=0$ and $q=0$ then $P_{++}=0$.  However, with $D>0$ and $q>0$ noise induced errors and decider interaction will ensure that $P_{++}>0$. 

\begin{figure}
\centering
  \includegraphics[width=0.9\textwidth]{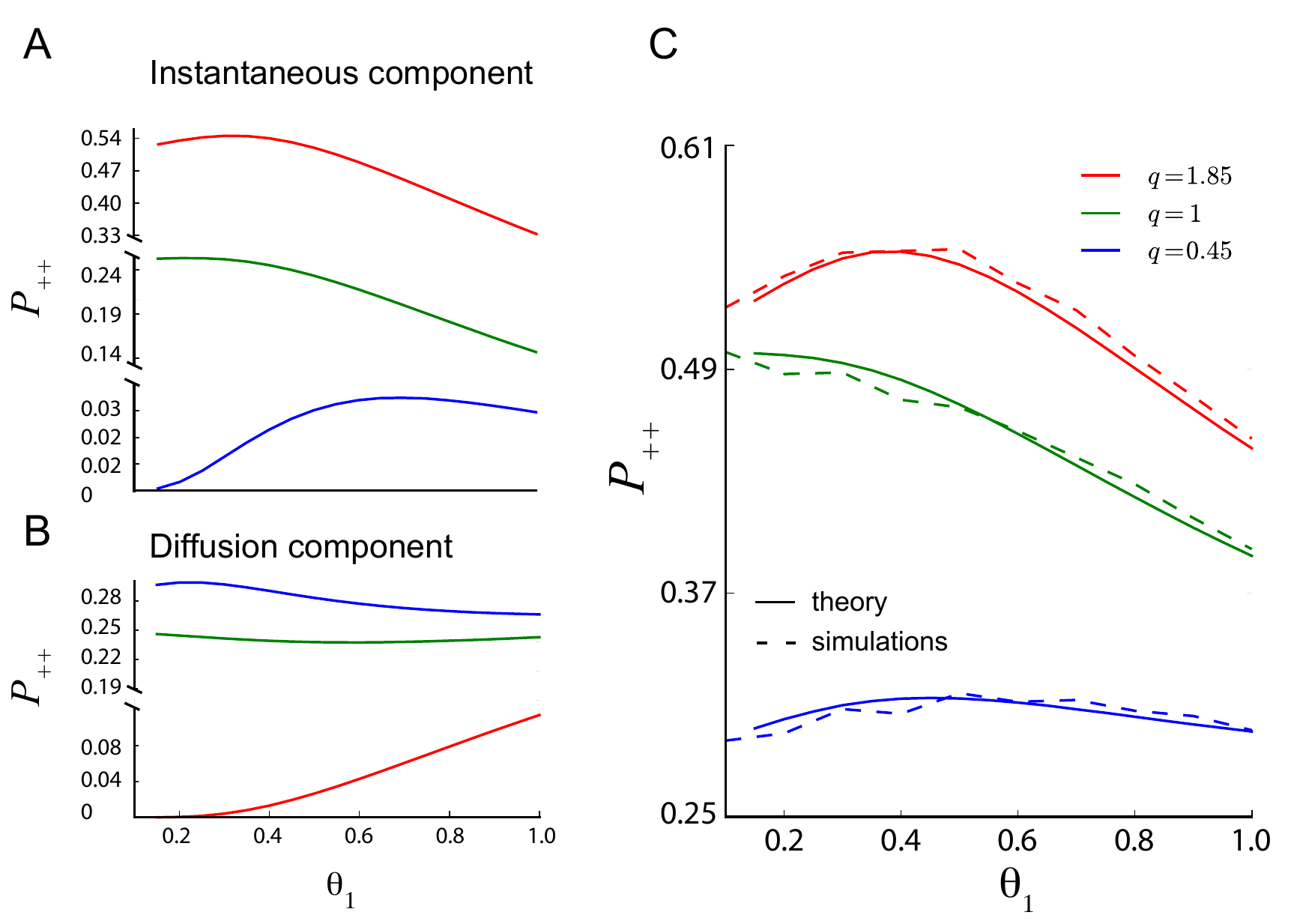}
    \caption{Instantaneous ({\bf A}) and diffusion ({\bf B}) components of $P_{++}$ for a two-decider system as $\theta_1$ varies.  Here $D=1$, $\mu_1=0.75$, $\mu_2=-0.75$, and $\theta_2=1$. {\bf C}. The solid lines are calculated directly from Eq. \eqref{eq:final}, while the dashed lines are calculated using simulations of Eqs. \eqref{MC1}-\eqref{MC2} with a stochastic Euler scheme ($\Delta t = 10^{-3}$, $10^5$ decisions for a fixed $\theta$).  In all panels we show results for the different interaction strengths ($q$) as indicated in the legend.}
    \label{Fig3}
\end{figure}

For various values of $q$ we consider the contributions to
$P_{++}$ that result from an instantaneous kick (Figure \ref{Fig3}A) and a kick with diffusion (Figure \ref{Fig3}B), as
well as their sum (Figure \ref{Fig3}C). 
When $q=0.45$ the instantaneous component is much smaller than the kick and
diffuse component (Figure \ref{Fig3}A,B, blue curves).  This is expected since the coupling is weak relative to the typical distance between $X_i$ and $\pm \theta_i$.  
In contrast, when $q=1.85$ the instantaneous component far outweighs the kick and diffuse component (Figure \ref{Fig3}A,B, red curves).  The total probability $P_{++}$ (sum of the two components) as derived from our analysis in Eq. \eqref{eq:final} gives a very accurate match to direct simulations of the decision processes described by the Langevin equations in Eqs. \eqref{MC1}-\eqref{MC2} (Figure \ref{Fig3}C, dashed vs. solid curves).

The central aim of our study is to understand how $\theta_1$ determines $P_{++}$.  For $q=0$ this is straightforward.  In this case the deciders are independent with $P(G_1=1,G_2=1)= P(G_1=1)P(G_2=1)$, and maximizing $P_{++}$ is equivalent to maximizing $P(G_1=1)$.  When $\mu_1>0$ then $P(G_1=1)$ only increases with $\theta_1$, since large decision thresholds are less susceptible to noise induced decision errors.  Thus, for groups without coupling decider 1 should have as high a decision threshold as possible, to at least be confident in their own decision.  Decision networks with $q>0$ give an interesting contrast to the independent case.  

For both small and large $q$, $P_{++}$ is maximized at a finite value of $\theta_1 < \theta_2=1$ (Figure \ref{Fig3}C, blue and red curves).  In other words, if a decider wishes to bias the group decision towards their personal bias then they should set their decision threshold to a lower value than if they were deciding in isolation from the group.
 While large $\theta_1$ mitigates the fluctuations in the decision process, it also forces the decision time $t_1$ to be large.  Recall that if $t_1 >t_2$ then decider 1 cannot influence decider 2, since decider 2 will have already decided.  Thus, in coupled networks there is a benefit to deciding early so as to influence the neighbor decider.  In this way the coupling introduces a form of 'speed accuracy tradeoff' in the group decision.

\begin{figure}

  \centering
 
  \includegraphics[width=0.9\textwidth]{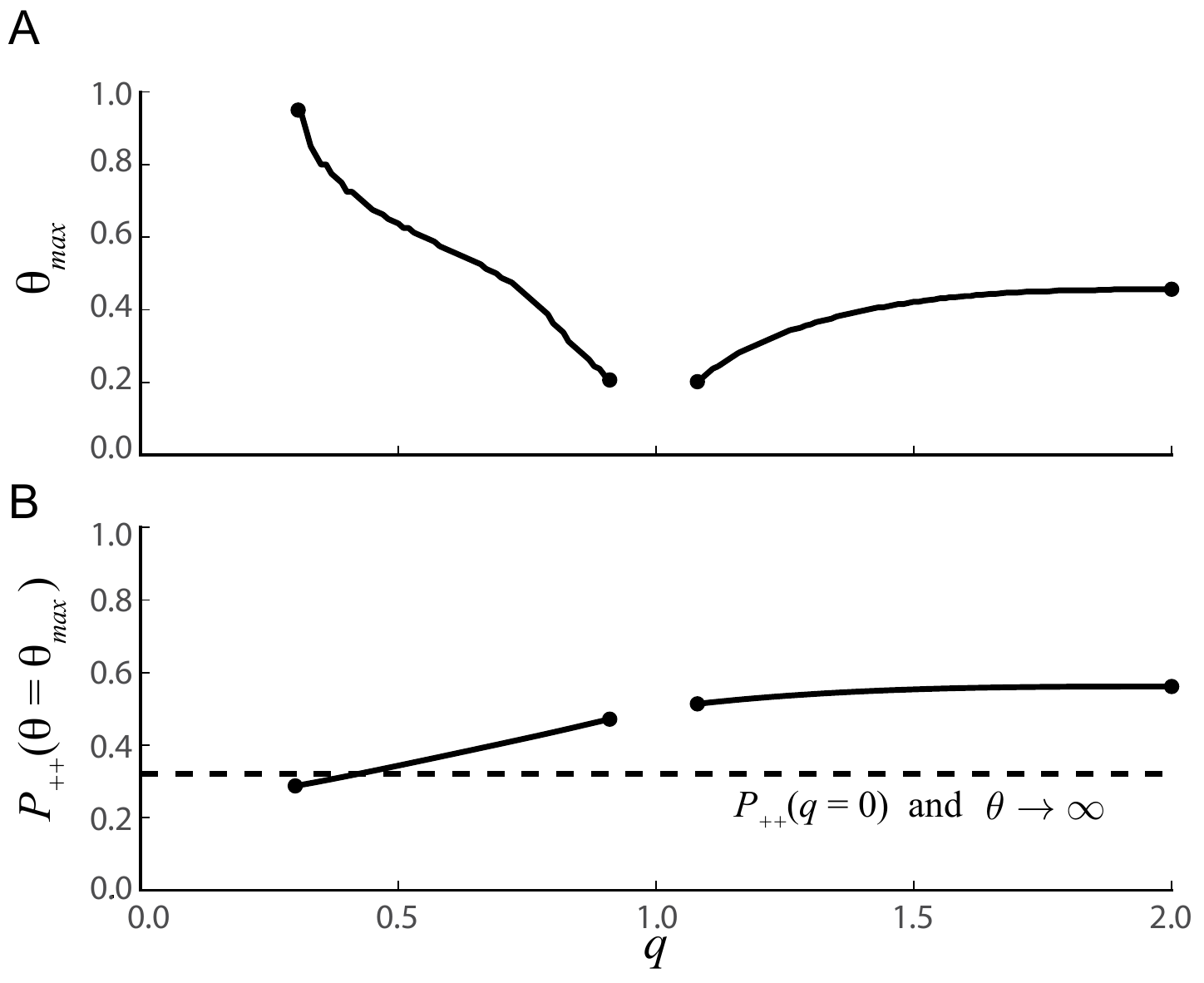}
    \caption{Values of $P_{++}$ and $\theta$ at the interior maximum as a function of the coupling $q$. The dashed line in the $P_{++}$ plot is the limiting value of the maximum $P_{++}$ when $q = 0$.}
    \label{Fig4}
\end{figure}

The interior maximum in $P_{++}$ as a function of $\theta_1$, labeled $\theta_{max}$, first appears at $q \approx 0.45$ (Figure~\ref{Fig4}).  As $q$ increases the value of $\theta_{max}$ decreases initially. This is because for larger $q$ the + decider has more influence on the $-$ decider, and to maximize $P_{++}$ through interaction it is best to have a lower value of $\theta_1$ so that there is a higher probability that $t_1 < t_2$. In a small region around $q=1$, the peak disappears, and then reappears for larger $q$. After it reappears, $\theta_{max}$ increases with $q$. This is because for large $q$ when the + decider makes its choice, it will likely immediately kick the other decider, forcing it to make the same choice. Thus, for $q>1$ it is best for it to have a higher threshold, thereby increasing the chance that decider 1 will cross the + threshold, which will likely result in the other decider making the + choice as well through an instaneous kick.  At the limit $q=2,$ the curves become saturated because when
one of the deciders makes its decision, it will always kick the other one over the same
threshold (since $q$ is twice the value of $\theta_2).$ 

In general, the value of $P_{++}$ at $\theta_{max}$ increases with $q$, since 
coupling will increase the probability of both making the same decision.
We remark that for large $q$ the group decision $P_{++}$ at $\theta_{max}$ is larger than the case for $q=0$ and $\theta \to \infty$ (Figure \ref{Fig4}B, dashed vs. solid). In other words, despite the $+$ decider losing accuracy with a lower decision threshold, with sufficient coupling the probability  of the $+,+$ group decision is higher than the optimal uncoupled case.

\begin{figure}[h!]
\centering
  \includegraphics[width=0.9\textwidth]{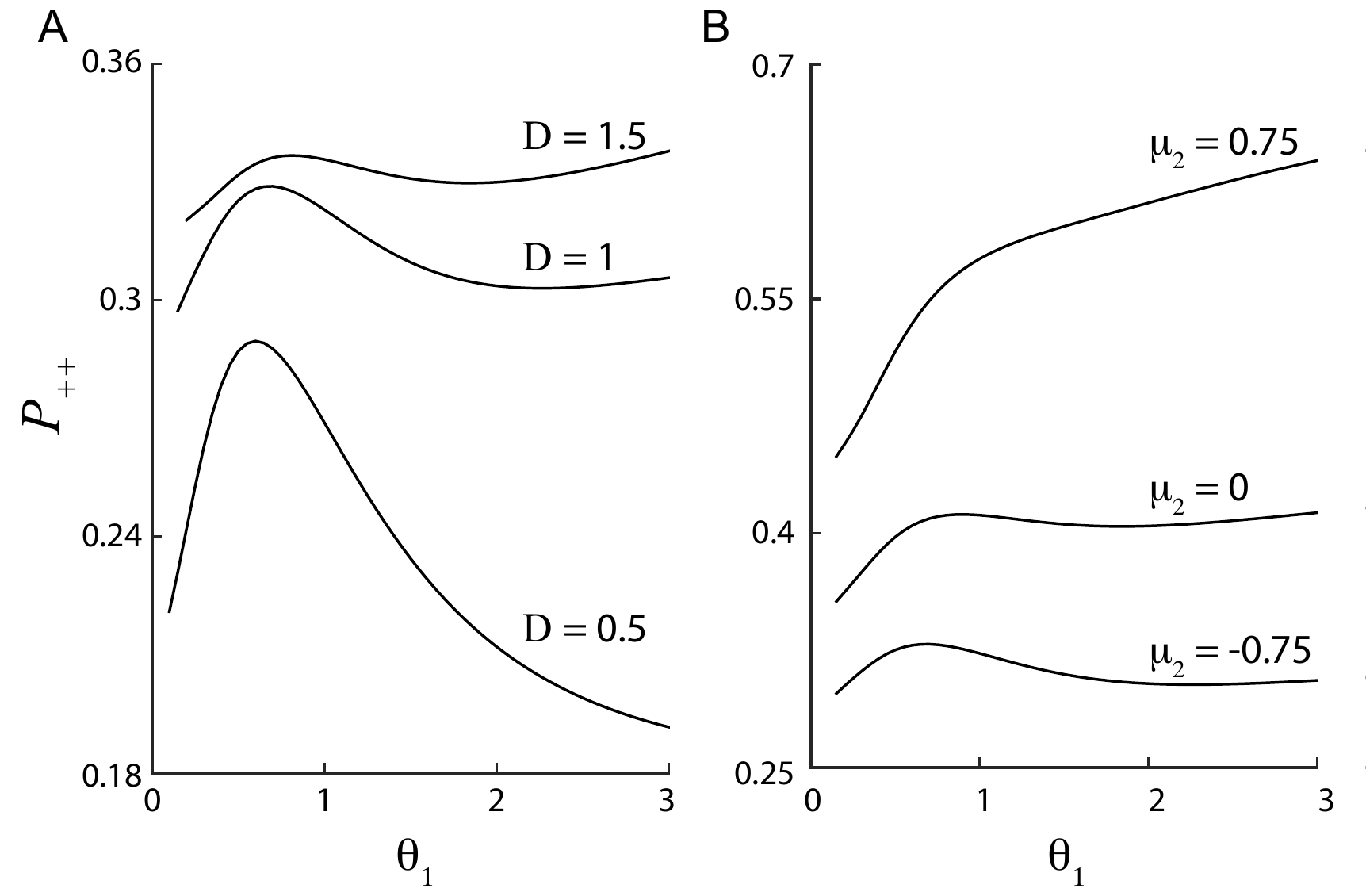}
    \caption{Joint probability of a +,+ group decision for various values of the diffusion coefficient $D$ ({\bf A}) and the bias for decider 2 $\mu_2$ ({\bf B}). Here $\mu_1=0.75$ and $q=0.45$.}
     \label{Fig_new}
\end{figure}

Finally, we asked whether the interior maximum in $P_{++}$ as a function of $\theta_1$ is a robust feature over a range in $\mu_2$ and $D$.  Overall, $P_{++}$ increases with $D$ since fluctuations are required for decider 2 (with $\mu_2=-0.75$) to cross the positive $\theta_2$ threshold.  For a wide range of $D$ a maximum occurs at a specific $\theta_1 < \theta_2=1$ (Figure \ref{Fig_new}A).  The maximum is also robust to changes in $\mu_2 <0$; however, the maximum disappears when $\mu_2$ becomes sufficiently positive (Figure \ref{Fig_new}B).  In this case decider 2 will have a tendency to cross the positive threshold even without coupling.  However, for $\theta_1 < \theta_2$ there is a larger gain in $P_{++}$ than for $\theta_1 > \theta_2$.  This is because for $\theta_1 < \theta_2$ the interaction can help overcome the fluctuations that causes errors in decider 2.  The robustness of a maximum in $P_{++}$ at a finite value of $\theta_1$ occurs for large $q$ as well (not shown).        

\section{Decision dynamics in larger groups}

\begin{figure}

  \centering
  \includegraphics[width=0.9\textwidth]{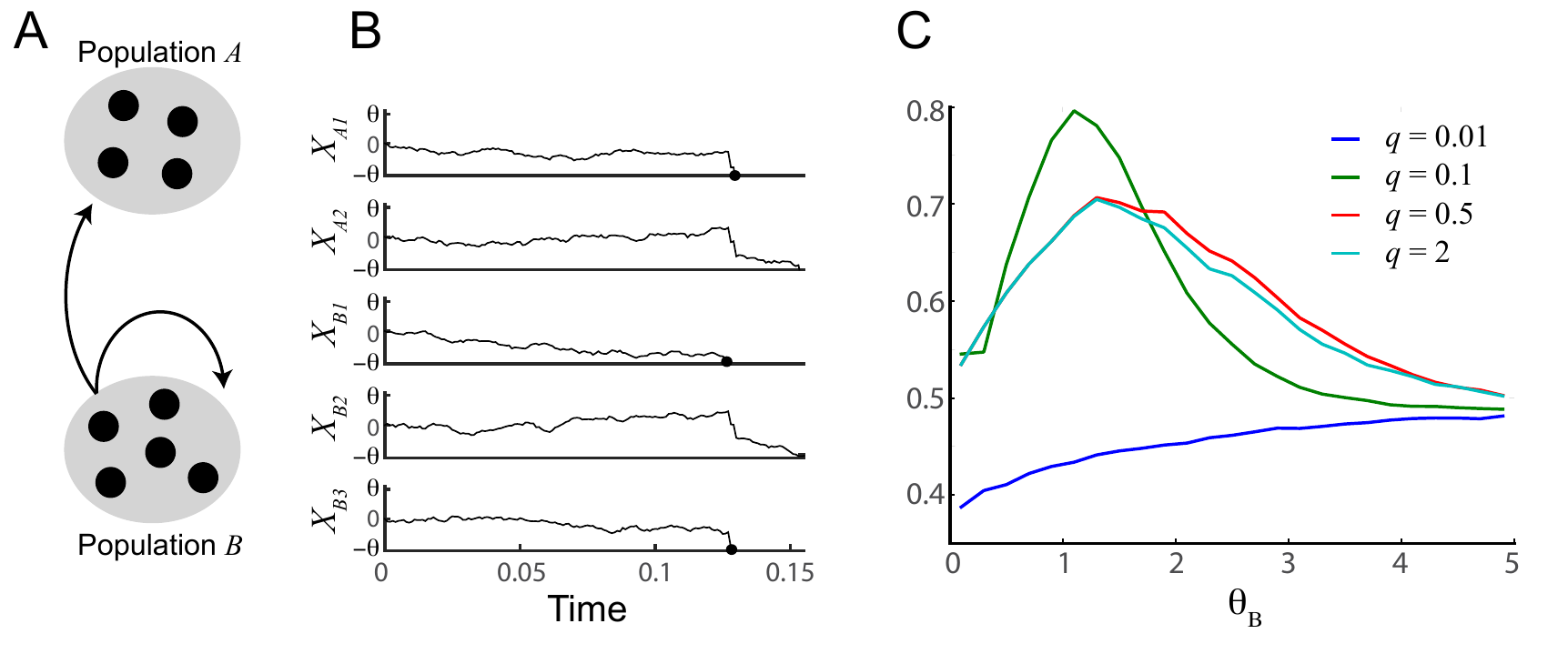}
    \caption{\textbf{A.} A schematic of the population. Members of B can influence other B members and A members, but A members can influence no one. \textbf{B.} A demonstration of the cascading effect discussed in the text. Populations A (2 members) and B (3 members) are as described in the text, with $Q=0.5$. Here, B1 escapes first, kicking all the other members down. This almost immediately pushes out B3 through the lower exit, further lowering A1, A2, and B2. \textbf{C.} Fraction $f_d$ of deciders that make the - choice.}
    \label{Fig5}
\end{figure}

Our analytic theory can be extended to the $N$-decider case (see Appendix). However, for $N>2$, the theory is cumbersome and we will simply explore the larger population case using numerical simulations.  We consider a total population of 100 deciders. We divide them into two populations, $A$ with $N_A=75$ deciders, and $B$ with $N_B=25$ deciders. The diffusion coefficient for all the deciders in the population is fixed at $D=1$. As in the two decider case the drift for deciders in group $A$ is $\mu= 0.75$, while the drift for deciders in $B$ is $-\mu$. Members of population $A$ have no influence on any member, so that $q_{AA}=q_{AB}=0,$ while members of $B$ influence everyone in the system with magnitude $q$, so that $q_{BA}=q_{BB}=q$ (Figure \ref{Fig5}A). The Langevin equations governing the evidence accumulation for this system are as follows. 

\begin{eqnarray}
dX_{Ai}&=&\mu dt+\sqrt{2D}dW_{Ai}+\sum\limits_{k=1}^{N_B} G_{Bk} q \delta(t-t_k), \\%
dX_{Bj}&=&-\mu dt+\sqrt{2D}dW_{Bj}+ \sum\limits_{k=1}^{N_B}G_{Bk} q \delta(t-t_k)(1-\delta_{jk}), \label{eq:popB}
\end{eqnarray}
where $1 \leq i \leq 75$, $1 \leq j \leq 25$, $t_k$ is the time at which decider $k$ in the B population decides, $G_{Bk}$ is $\pm 1$ if decider $k$ chose the $\pm$ decision.  The $1-\delta_{ik}$ term in Eq. \eqref{eq:popB} removes self coupling within population $B$.   Sample realizations show stochastic decision dynamics similar to the two decider case (Figure \ref{Fig5}B).  

Let $f_d$ be the fraction of deciders that choose the $-$ decision. For $q>0$ this measures the influence that the smaller, interacting part of the population has on the group as a whole; i.e. the more deciders that choose the $-$ decision, the more influence population $B$ has on the total population.  In the limit $q \rightarrow 0$, to maximize $f_d$ we should send $\theta_B \rightarrow \infty$ to maximize the chance that $B$ deciders cross through the negative threshold. However, as $q$ increases $f_d$ is maximized at a finite value of $\theta_B$ (Figure \ref{Fig5}C; $\theta_A=1$).  This maximum can be as high as $f_d \approx 0.8$ -- a quarter of the population has made 80 percent of the total members of the population cross through the $-$ threshold towards which the the 25 members of population $B$ are biased (Figure~\ref{Fig5}C, $q=0.1$). 

Note that, unlike in the two-decider case discussed above, the value of the peak first increases with $q$, but then it decreases and eventually saturates (Figure \ref{Fig5}C, $q=0.5$ and $q=2$ curves). This can be explained as follows. When the first $B$ decider makes its decision, it will make the $\pm$ choice and will shift the evidences of all of the other deciders in the $\pm$ direction. If $q$ is large, this will cause a large number of deciders (some of which will, of course, be $B$ deciders) to cross through the $\pm$ threshold. These will in turn shift the remaining deciders, causing some of them to cross, and so on and so forth. In this way, for large enough values of $q$, we have a cascading effect--the fate of almost all the deciders is determined by a single (uncertain) decider.  This dynamic is akin to herd behavior where members of the group are driven primarily by neighbor decisions rather than their own evidence accumulation \cite{banerjee1992}.
 
On the other hand, for smaller values of $q$ one decider will have some influence on the system, but not enough to solely influence a large fraction of the population.  Rather, the system's overall behavior is dependent on more than one decider and is thus subject to less noise. This means that a higher fraction of the deciders will decide on the choice towards which $B$ is biased. 

Our simulations of the $N$-decider system show that larger populations qualitatively match the key feature of the 2-decider case.  Namely, for nonzero coupling it is possible to choose a decision threshold so as to maximize the probability of a coordinated group decision.  

\section{Discussion}

In this paper we modeled interactions between two or more TAFC drift-diffusion models. In the two-decider model the deciders were biased towards opposite decisions. By varying the threshold for the + decider, we determined how $P_{++}$ (the probability that both deciders choose the + gate) is controlled for various values of the coupling strength $q$. When $q=0$, i.e., no coupling, $P_{++}$ is maximized for decision threshold $\theta \to \infty$.  Our main finding is that for a large range of $q>0$, $P_{++}$ is maximized at a finite $\theta$ for the + decider.  The intuition for the two decider case was extended to a large population of coupled deciders, where we demonstrated that with an appropriately chosen threshold a small but strongly coupled subgroup of the population can have a large impact on the group decision.  

In developing the theory of the two-decider system, we had to consider integrals over one of the time variables (see Eq. \eqref{eq:cFPT}).  In principle, this can be extended to the case with $N$ deciders. However, in order to complete this, we need to analyze the combinatorics of the orders in which the deciders make their decisions (see Appendix). The analog of \eqref{eq:cFPT} for the $N$-decider case will be an $N-1$ dimensional integral (see Eq. \eqref{eq:N_dim_int}). Evaluating a large number of these high-dimensional integrals is computationally cumbersome.

Previous studies have considered the collective decision making of groups of drift-diffusion models.  One class of model considers the accumulation dynamics in populations of uncoupled agents with a threshold decision rule (same as Eqs.~\eqref{MC1}-\eqref{MC2} with $q=0$) combined with a consensus group decision \cite{dandach2012,kimura2012}.  In such models the independent diffusion permits a clear analysis to be performed.  Another class of model considers  evidence accumulation in a population where deciders linearly couple their evidence \cite{srivastava2014,poulakakis2010}.  Here the interaction is continuous in time, yet the decision mechanics do not involve thresholded evidence, rather the accumulation is free running.  The linearity of the model permits an analysis of the full population accumulation.    Our framework is distinct from these models in that it combines both interaction during the accumulation process and a threshold decision rule.  Evidence accumulation is only shared at decision times, and otherwise accumulation is independent between deciders, permitting an analysis of group activity (Eq. \eqref{eq:final}).  However, both the uncoupled population model with consensus and the linear free running population model have analysis that scales well with system size, unlike our model.

Our model exhibits a form of speed-accuracy tradeoff \cite{wickelgren1977,bogacz2010,bogacz2010n,bogacz2006,srivastava2014}. In the TAFC task there is a tradeoff in the decision making process: the decider would like to make the correct decision in the shortest amount of time. To increase the probability that it makes the correct choice the decider can increase the amount of evidence that it requires to make a decision (i.e., the threshold $\theta$). However, larger decision thresholds increase the amount of time it takes to reach the threshold, and hence decision accuracy and decision speed are at odds with one another. 

In the classical speed-accuracy tradeoff for a single TAFC drift-diffusion model, the reward for making a correct decision quickly is expressed via a `cost function' that is added to the model by hand \cite{bogacz2006}. However, in our group decision model the speed-accuracy tradeoff emerges from the group interaction. The decision agents still want to be accurate, but the incentive for speed is not a built-in cost function. Rather, the reward for an individual decider to make its choice quickly (and correctly) is the chance for it to have influence other members of the population, and thus increase the fraction of population members that decide its correct choice.

The single decider drift diffusion model has been a very influential in decision theory \cite{Vickers1970,ratcliff1978,usher2001,bogacz2006,ratcliff2008}, in large part because one can compute the threshold value $\theta$ that will optimize the reward rate received by the decider \cite{bogacz2006}.  Combining this theory and efficient statistical techniques to estimate the drift and diffusion terms from a collection of decision experiments gives a prescription to test whether deciders are acting optimally.  Our model provides a theory for the decision outcome and times from groups of coupled deciders.  It remains to extend our theory in Eq. \eqref{eq:final} to compute optimal thresholds ($\theta$) and interactions ($q$) under a set of task constraints.  With this in hand it may be possible to determine whether groups of deciders act optimally.          


\clearpage
\appendix
\section{Theory for $N$-Decider case}
Suppose we have $N$ deciders where decider $i$ has evidence accumulation described by $D_i,$ $\mu_i,$ and  $\theta_i$. A decider pair has a
coupling $q_{ij}$ representing the influence of $j$ on $i.$ If decider
$i$ escapes through the threshold $\pm\theta_{i}$, we assign the value $\pm1$
to $G_{i}$ ($1\leq i\leq N)$. We wish to calculate:
\[
P\left(  G_{1},\ldots,G_{N},\text{ord}\right),
\]
where ord refers to the decision order under consideration; without loss of generality we take $\text{ord}=t_{1}%
<t_{2}<\cdots<t_{N}.$ This probability can be written as follows
\[
P\left(  G_{1},\ldots,G_{N},\text{ord}\right)  =\int_{0}^{\infty}%
dt_{1}P\left(  G_{1},t_{1},\text{ord}\right)  P\left(  G_{2},\ldots
,G_{N}|G_{1},t_{1},\text{ord}\right)  .
\]

We have that:

\begin{equation}
P\left(  G_{1},t_{1},\text{ord}\right) =\int_{\text{ord}}f_{1}^{G_{1}}\left(  t_{1}\right)
{\displaystyle\prod\limits_{k=2}^{N}}
f_{k}\left(  t_{k}\right)  dt_{2}\cdots dt_{N}.%
\label{eq:N_dim_int}
\end{equation}

Define $f_{1,\text{ord}}^{G_{1}}\left(  t_{1}\right)  \equiv\int_{\text{ord}%
}f_{1}^{G_{1}}\left(  t_{1}\right)
{\displaystyle\prod\limits_{k=2}^{N}}
f_{k}\left(  t_{k}\right)  dt_{2}\cdots dt_{N}$ so that
\[
P\left(  G_{1},\ldots,G_{N},\text{ord}\right)  =\int_{0}^{\infty}%
dt_{1}f_{1,\text{ord}}^{G_{1}}\left(  t_{1}\right)  P\left(  G_{2}%
,\ldots,G_{N}|G_{1},t_{1},\text{ord}\right)  .
\]

We now wish to calculate $P\left(  G_{2},\ldots,G_{N}|G_{1},t_{1}%
,\text{ord}\right)$. Note that decider $2$ makes its decision after decider $1$ but before decider 3, and so on. At each stage there are two possibilities: either the decider is kicked
instantly across the desired threshold (which is only possible if
$G_{i}=G_{i-1}$), or it diffuses and escapes later. Now, define
the probability density ($t_{j}$ is the time at which decider $j$ makes its decision) as:\
\[
\rho_{i}\left(  x_i,t;G_{1}q_{i1},t_{1},\ldots,G_{k}q_{ik},t_{k}\right)
\]
to be the density of decider $i$'s evidence at time $t$ after the kicks
$G_{1}q_{i1}$ at $t_{1},\ldots,G_{k}q_{ik},$ at time $t_{k}.$ Define
$\rho_{i}\left(  x_j,t;k\right)  \equiv\rho_{i}\left(  x_i,t;G_{1}q_{i1}%
,t_{1},\ldots,G_{k}q_{ik},t_{k}\right).$

If decider $1$ escapes at $t_{1},$ decider $2$ can escape instantly so that $t_{2}%
=t_{1},$ or it can escape at some later time $t_{2}>t_{1}.$ In either of these
situations, it can kick $3$ across instantly, or some time later, and so on,
until the $N$th decider. This can be seen below schematically.%

\[
\text{decider 1 crosses at }t_{1}\longrightarrow\Biggl\lbrace\
\begin{array}
[c]{c}%
2\text{ kicked inst.}\longrightarrow\Biggl\lbrace\
\begin{array}
[c]{c}%
3\text{ kicked inst.}\longrightarrow\cdots

\\
3\text{ crosses later}\longrightarrow\cdots

\end{array}
\\
2\text{ crosses later}\longrightarrow\Biggl\lbrace\
\begin{array}
[c]{c}%
3\text{ kicked inst.}\longrightarrow\cdots

\\
3\text{ crosses later}\longrightarrow\cdots

\end{array}
\end{array}
\]

\bigskip For each $2\leq i\leq N$ define the variable $k_{i}$ to be
\[
k_{i}=\Biggl\lbrace\
\begin{array}
[c]{c}%
0,\text{ }i\text{ kicked instantly after }i-1\\
1,\text{ }i\text{ escapes sometime later through correct gate}%
\end{array}
\]

Now, consider the decision of the $i$th decider. Suppose that the previous
deciders made their decisions at times $t_{1},\ldots,t_{k}.$ If $k_{i}=0,$ the
probability that decider makes an instantaneous threshold crossing is:
\[
\int_{\ell_{i}^{i-1}}^{u_{i}^{i-1}}\rho_{i}\left(  x_i,t_{i-1};i-2\right)  dx
\]
where
\[
\ell_{i}^{i-1}=\ell_{i}^{i-1}\left(  G_{i},G_{i-1}\right)  =\Biggl\lbrace\
\begin{array}
[c]{c}%
\theta_{i}-G_{i-1}q_{i}^{i-1},\text{ }G_{i}>0\\
-\theta_{i},\text{ }G_{i}<0
\end{array}
\]

\[
u_{i}^{i-1}=u_{i}^{i-1}\left(  G_{i},G_{i-1}\right)  =\Biggl\lbrace\
\begin{array}
[c]{c}%
\theta_{i},\text{ }G_{i}>0\\
-\theta_{i}-G_{i-1}q_{i}^{i-1},\text{ }G_{i}<0.
\end{array}
.
\]

The probability that the decision is not made instantly is:%
\[
\int_{L_{i}^{i-1}}^{U_{i}^{i-1}}\rho_{i}\left(  x_i,t_{i-1};i-2\right)  dx
\]

with
\[
L_{i}^{i-1}=L_{i}^{i-1}\left(  G_{i},G_{i-1}\right)  =\Biggl\lbrace\
\begin{array}
[c]{c}%
-\theta,\text{ }G_{i-1}>0\\
-\theta_{i}+q_{i}^{i-1},\text{ }G_{i-1}<0
\end{array}
\]

\[
U_{i}^{i-1}=U_{i}^{i-1}\left(  G_{i},G_{i-1}\right)  =\Biggl\lbrace\
\begin{array}
[c]{c}%
\theta_{i}-q_{i}^{i-1},\text{ }G_{i-1}>0\\
\theta_{i},\text{ }G_{i-1}<0
\end{array}
.
\]

Given that decider $i$ does not decide instantly after $i-1$, the first passage time
density $f_{i}^{G_{i}}\left(  t_{i};i-1\right)  $ is obtained by calculating
the flux from $\rho_{i}\left(  x_i,t_{i};i-1\right)  ,$ normalized such that
\[
\int_{t_{i-1}}^{\infty}\left[  f_{i}^{G_{i}}\left(  t_{i};i-1\right)
+f_{i}^{-G_{i}}\left(  t_{i};i-1\right)  \right]  dt_{i}=1.
\]
Thus the contribution from this case is:
\[
\int_{L_{i}^{i-1}}^{U_{i}^{i-1}}\rho_{i}\left(  x_i,t_{i-1};i-2\right)
dx\int_{t_{i-1}}^{\infty}f_{i}^{G_{i}}\left(  t_{i};i-1\right)dt_{i}.
\]
Note that everything after decider $i$ depends on $t_{i}$ so we are
integrating these over $t_{i}$ as well.

Now, at each stage (i.e. for each value of $i>1$), $k_{i}$ can be
$0$ or $1.$ Summing over all possible values of the
$k_{i}$'s gives us all cases for a given set of decisions ($G$'s) and a time ordering:
\begin{align*}
P\left(  G_{1},\ldots,G_{N},\text{ord}\right)   &  =\int_{0}^{\infty}%
f_{1,\text{ord}}^{G_{1}}\left(  t_{1}\right)  P\left(  G_{2}%
,\ldots,G_{N}|G_{1},t_{1},\text{ord}\right)dt_{1} \\
&  =\sum_{k_{2},\ldots,k_{n}}\int_{0}^{\infty}f_{1,\text{ord}}^{G_{1}%
}\left(  t_{1}\right)  P\left(  G_{2},\ldots,G_{N},k_{2},\ldots,k_{n}%
|G_{1},t_{1},\text{ord}\right)dt_{1}
\end{align*}

For a given set of values for $k_{2},\ldots,k_{i-1},$ the contribution from
decider $i$ is
\[
\mathcal{C}_{i}\left(  k_{i};k_{2},\ldots,k_{i-1}\right)  =\Biggl\lbrace\
\begin{array}
[c]{c}%
\int_{\ell_{i}^{i-1}}^{u_{i}^{i-1}}\rho_{i}\left(  x_i,t_{i-1};i-2\right)
dx,\text{ }k_{i}=0\\
\int_{L_{i}^{i-1}}^{U_{i}^{i-1}}\rho_{i}\left(  x_i,t_{i-1};i-2\right)
dx\int_{t_{i-1}}^{\infty}f_{i}^{G_{i}}\left(  t_{i};i-1\right)dt_{i}  ,\text{
}k_{i}=1.
\end{array}
\]
As well,
\[
P\left(  G_{2},\ldots,G_{N},k_{2},\ldots,k_{n}|G_{1},t_{1},\text{ord}\right)
=%
{\displaystyle\prod\limits_{i=2}^{N}}
\mathcal{C}_{i}\left(  k_{i};k_{2},\ldots,k_{i-1}\right)
\]
so that
\[
P\left(  G_{2},\ldots,G_{N},\text{ord}\right)  =\sum_{k_{2},\ldots,k_{n}}%
\int_{0}^{\infty}f_{1,\text{ord}}^{G_{1}}\left(  t_{1}\right)dt_{1}
{\displaystyle\prod\limits_{i=2}^{N}}
\mathcal{C}_{i}\left(  k_{i};k_{2},\ldots,k_{i-1}\right)  .
\]

We can define $\mathcal{C}_{1}\equiv\int_{0}^{\infty}dt_{1}f_{1,\text{ord}%
}^{G_{1}}\left(  t_{1}\right)  $ so that
\[
P\left(  G_{2},\ldots,G_{N},\text{ord}\right)  =\sum_{k_{2},\ldots,k_{n}}%
{\displaystyle\prod\limits_{i=1}^{N}}
\mathcal{C}_{i}\left(  k_{i};k_{2},\ldots,k_{i-1}\right)  .
\]

As an application consider a system with three deciders; two are in
the $A$ population, and one is in the $B$ population. Let $D_{a}=D_{b}=D,$ and
$\mu_{B}=-\mu_{A}=-\mu.$ The deciders in the $A$ population ($A1$ and $A2$)
influence the other system members with magnitude $q_{A}$ and the $B$ decider
influences the other deciders with magnitude $q_{B}.$
Up to permutations of the $A$ deciders, there are 3 ways of ordering the
deciders:\
\begin{align*}
\text{ord 1}\text{: }  &  t_{a1}<t_{a2}<t_{b}\\
\text{ord 2}\text{: }  &  t_{a1}<t_{b}<t_{a2}\\
\text{ord 3}\text{: }  &  t_{b}<t_{a1}<t_{a2}.
\end{align*}

Let us calculate $P\left(  G_{1}%
,G_{2},G_{3},\text{ord 1}\right)$ (labeling the deciders as $1\longleftrightarrow A1,$
$2\longleftrightarrow A2,$ and $3\longleftrightarrow B$). Using the scheme outlined above,
\[
A1\text{ crosses at }t_{A1}\longrightarrow\Biggl\lbrace\
\begin{array}
[c]{c}%
k_{2}=0\longrightarrow\Biggl\lbrace\
\begin{array}
[c]{c}%
k_{3}=0\\
k_{3}=1
\end{array}
\\
k_{2}=1\longrightarrow\Biggl\lbrace\
\begin{array}
[c]{c}%
k_{3}=0\\
k_{3}=1
\end{array}
\end{array}
\]

So, using the formulas for each case,
\begin{align*}
&  P\left(  G_{1},G_{2},G_{3},\text{ord 1}\right) \\
&  =\int_{0}^{\infty}f_{A,\text{ord1}}^{G_{1}}\left(  t_{1}\right)dt_{1}
\left\{
\begin{array}
[c]{c}%
\delta_{G_{1}G_{2}}\int_{\ell_{2}^{1}}^{u_{2}^{1}}\rho_{A}\left(
x,t_{1}\right)  dx \\ \left[
\begin{array}
[c]{c}%
\delta_{G_{2}G_{3}}\int_{\ell_{3}^{2}}^{u_{3}^{2}}\rho_{B}\left(
x,t_{1};1\right)  dx\\
+\int_{L_{3}^{2}}^{U_{3}^{2}}\rho_{B}\left(  x,t_{1};t_{1}\right)
dx\int_{-\theta_{B}}^{\theta_{B}}\rho_{B}\left(  x,t_{1};t_{1},t_{1}\right)
\varepsilon_{B}^{G_{3}}\left(  x\right)  dx
\end{array}
\right] \\
+\int_{L_{2}^{1}}^{U_{2}^{1}}\rho_{A}\left(  x,t_{1}\right)  dx\int_{t_{1}%
}^{\infty}dt_{2}f_{a}^{G_{2}}\left(  t_{2};1\right)  \\ \left[
\begin{array}
[c]{c}%
\delta_{G_{2}G_{3}}\int_{\ell_{3}^{2}}^{u_{3}^{2}}\rho_{B}\left(
x,t_{2};1\right)  dx\\
+\int_{L_{3}^{2}}^{U_{3}^{2}}\rho_{B}\left(  x,t_{2};t_{1}\right)
dx\int_{-\theta_{B}}^{\theta_{B}}\rho_{B}\left(  x,t_{2};t_{1},t_{2}\right)
\varepsilon_{B}^{G_{3}}\left(  x\right)  dx
\end{array}
\right]
\end{array}
\right\}
\end{align*}
$\ $

To get ord 2 from ord 1, we need to swap decider $2$ with decider $3.$%
\begin{align*}
&  P\left(  G_{1},G_{2},G_{3},\text{ord 2}\right) \\
&  =\int_{0}^{\infty}f_{A,\text{ord2}}^{G_{1}}\left(  t_{1}\right)dt_{1}
\left\{
\begin{array}
[c]{c}%
\delta_{G_{1}G_{3}}\int_{\ell_{3}^{1}}^{u_{3}^{1}}\rho_{B}\left(
x,t_{1}\right)  dx\left[
\begin{array}
[c]{c}%
\delta_{G_{2}G_{3}}\int_{\ell_{2}^{3}}^{u_{2}^{3}}\rho_{A}\left(
x,t_{1};1\right)  dx\\
+\int_{L_{2}^{3}}^{U_{2}^{3}}\rho_{a}\left(  x,t_{1};t_{1}\right)
dx\int_{-\theta_{a}}^{\theta_{A}}\rho_{A}\left(  x,t_{1};t_{1},t_{1}\right)
\varepsilon_{A}^{G_{2}}\left(  x\right)  dx
\end{array}
\right] \\
+\int_{L_{3}^{1}}^{U_{3}^{1}}\rho_{B}\left(  x,t_{1}\right)  dx\int_{t_{1}%
}^{\infty}f_{B}^{G_{3}}\left(  t_{3};1\right)dt_{3}  \\ \left[
\begin{array}
[c]{c}%
\delta_{G_{2}G_{3}}\int_{\ell_{2}^{3}}^{u_{2}^{3}}\rho_{A}\left(
x,t_{3};1\right)  dx\\
+\int_{L_{2}^{3}}^{U_{2}^{3}}\rho_{a}\left(  x,t_{3};t_{1}\right)
dx\int_{-\theta_{A}}^{\theta_{A}}\rho_{A}\left(  x,t_{3};t_{1},t_{3}\right)
\varepsilon_{A}^{G_{2}}\left(  x\right)  dx
\end{array}
\right]
\end{array}
\right\}
\end{align*}
To get ord 3 from ord 2, we switch decider $1$ with decider $3.$%
\begin{align*}
&  P\left(  G_{1},G_{2},G_{3},\text{ord 3}\right) \\
&  =\int_{0}^{\infty}f_{B,\text{ord3}}^{G_{3}}\left(  t_{3}\right)dt_{3}
\left\{
\begin{array}
[c]{c}%
\delta_{G_{1}G_{3}}\int_{\ell_{1}^{3}}^{u_{1}^{3}}\rho_{A}\left(
x,t_{3}\right)  dx\left[
\begin{array}
[c]{c}%
\delta_{G_{2}G_{1}}\int_{\ell_{2}^{1}}^{u_{2}^{1}}\rho_{A}\left(
x,t_{3};1\right)  dx\\
+\int_{L_{2}^{1}}^{U_{2}^{1}}\rho_{A}\left(  x,t_{3};t_{3}\right)
dx\int_{-\theta_{A}}^{\theta_{A}}\rho_{A}\left(  x,t_{3};t_{3},t_{3}\right)
\varepsilon_{A}^{G_{2}}\left(  x\right)  dx
\end{array}
\right] \\
+\int_{L_{1}^{3}}^{U_{1}^{3}}\rho_{A}\left(  x,t_{3}\right)  dx\int_{t_{3}%
}^{\infty}f_{A}^{G_{1}}\left(  t_{1};1\right)dt_{1}  \\ \left[
\begin{array}
[c]{c}%
\delta_{G_{2}G_{1}}\int_{\ell_{2}^{1}}^{u_{2}^{1}}\rho_{A}\left(
x,t_{3};1\right)  dx\\
+\int_{L_{2}^{1}}^{U_{2}^{1}}\rho_{A}\left(  x,t_{3};t_{1}\right)
dx\int_{-\theta_{a}}^{\theta_{A}}\rho_{A}\left(  x,t_{3};t_{3},t_{1}\right)
\varepsilon_{A}^{G_{2}}\left(  x\right)  dx
\end{array}
\right]
\end{array}
\right\}
\end{align*}
Here $\rho_{A}\left(  x,t;1\right)  $ represents the density of $A$
after the first kick, and $\rho_{B}\left(  x,t;t_{1},t_{3}\right)  $ is the
density of $B$ after the first kick at $t_{1}$ and the second kick at $t_{3},
$ and $\varepsilon_{i}^{\pm1}\left(  x\right)  $ is the probability that a
decider of type $i$ $\left(  =A\text{ or }B\right)  $ at some evidence $x$ will
cross threshold $\pm\theta_{i}$. Now, for each of the ord 1, ord 2,
ord 3, we can swap the two $A$ deciders and get the same answer; these
represent all possible time orderings. Hence,
\begin{align*}
&  P\left(  G_{1},G_{2},G_{3}\right) \\
&  =2\sum_{i=1}^{3}P\left(  G_{1},G_{2},G_{3},\text{ord }i\right) \\
&  =2\int_{0}^{\infty}f_{A,\text{ord1}}^{G_{1}}\left(  t_{1}\right)dt_{1}
\left\{
\begin{array}
[c]{c}%
\delta_{G_{1}G_{2}}\int_{\ell_{2}^{1}}^{u_{2}^{1}}\rho_{A}\left(
x,t_{1}\right)  dx\left[
\begin{array}
[c]{c}%
\delta_{G_{2}G_{3}}\int_{\ell_{3}^{2}}^{u_{3}^{2}}\rho_{B}\left(
x,t_{1};1\right)  dx\\
+\int_{L_{3}^{2}}^{U_{3}^{2}}\rho_{B}\left(  x,t_{1};t_{1}\right)
dx\int_{-\theta_{B}}^{\theta_{B}}\rho_{B}\left(  x,t_{1};t_{1},t_{1}\right)
\varepsilon_{B}^{G_{3}}\left(  x\right)  dx
\end{array}
\right] \\
+\int_{L_{2}^{1}}^{U_{2}^{1}}\rho_{A}\left(  x,t_{1}\right)  dx\int_{t_{1}%
}^{\infty}dt_{2}f_{A}^{G_{2}}\left(  t_{2};1\right) \\  \left[
\begin{array}
[c]{c}%
\delta_{G_{2}G_{3}}\int_{\ell_{3}^{2}}^{u_{3}^{2}}\rho_{B}\left(
x,t_{2};1\right)  dx\\
+\int_{L_{3}^{2}}^{U_{3}^{2}}\rho_{B}\left(  x,t_{2};t_{1}\right)
dx\int_{-\theta_{B}}^{\theta_{B}}\rho_{B}\left(  x,t_{2};t_{1},t_{2}\right)
\varepsilon_{B}^{G_{3}}\left(  x\right)  dx
\end{array}
\right]
\end{array}
\right\} \\
&  +2\int_{0}^{\infty}dt_{1}f_{A,\text{ord2}}^{G_{2}}\left(  t_{1}\right)
\left\{
\begin{array}
[c]{c}%
\delta_{G_{1}G_{3}}\int_{\ell_{3}^{1}}^{u_{3}^{1}}\rho_{B}\left(
x,t_{1}\right)  dx\left[
\begin{array}
[c]{c}%
\delta_{G_{2}G_{3}}\int_{\ell_{2}^{3}}^{u_{2}^{3}}\rho_{A}\left(
x,t_{1};1\right)  dx\\
+\int_{L_{2}^{3}}^{U_{2}^{3}}\rho_{A}\left(  x,t_{1};t_{1}\right)
dx\int_{-\theta_{A}}^{\theta_{A}}\rho_{A}\left(  x,t_{1};t_{1},t_{1}\right)
\varepsilon_{A}^{G_{2}}\left(  x\right)  dx
\end{array}
\right] \\
+\int_{L_{3}^{1}}^{U_{3}^{1}}\rho_{B}\left(  x,t_{1}\right)  dx\int_{t_{1}%
}^{\infty}dt_{3}f_{B}^{G_{3}}\left(  t_{3};1\right) \\  \left[
\begin{array}
[c]{c}%
\delta_{G_{2}G_{3}}\int_{\ell_{2}^{3}}^{u_{2}^{3}}\rho_{A}\left(
x,t_{3};1\right)  dx\\
+\int_{L_{2}^{3}}^{U_{2}^{3}}\rho_{A}\left(  x,t_{3};t_{1}\right)
dx\int_{-\theta_{A}}^{\theta_{A}}\rho_{A}\left(  x,t_{3};t_{1},t_{3}\right)
\varepsilon_{A}^{G_{2}}\left(  x\right)  dx
\end{array}
\right]
\end{array}
\right\} \\
&  +2\int_{0}^{\infty}dt_{3}f_{B,\text{ord3}}^{G_{3}}\left(  t_{3}\right)
\left\{
\begin{array}
[c]{c}%
\delta_{G_{1}G_{3}}\int_{\ell_{1}^{3}}^{u_{1}^{3}}\rho_{A}\left(
x,t_{3}\right)  dx\left[
\begin{array}
[c]{c}%
\delta_{G_{2}G_{1}}\int_{\ell_{2}^{1}}^{u_{2}^{1}}\rho_{A}\left(
x,t_{3};1\right)  dx\\
+\int_{L_{2}^{1}}^{U_{2}^{1}}\rho_{A}\left(  x,t_{3};t_{3}\right)
dx\int_{-\theta_{A}}^{\theta_{A}}\rho_{A}\left(  x,t_{3};t_{3},t_{3}\right)
\varepsilon_{A}^{G_{2}}\left(  x\right)  dx
\end{array}
\right] \\
+\int_{L_{1}^{3}}^{U_{1}^{3}}\rho_{A}\left(  x,t_{3}\right)  dx\int_{t_{3}%
}^{\infty}dt_{1}f_{A}^{G_{1}}\left(  t_{1};1\right)   \\ \left[
\begin{array}
[c]{c}%
\delta_{G_{2}G_{1}}\int_{\ell_{2}^{1}}^{u_{2}^{1}}\rho_{A}\left(
x,t_{3};1\right)  dx\\
+\int_{L_{2}^{1}}^{U_{2}^{1}}\rho_{A}\left(  x,t_{3};t_{1}\right)
dx\int_{-\theta_{A}}^{\theta_{A}}\rho_{A}\left(  x,t_{3};t_{3},t_{1}\right)
\varepsilon_{A}^{G_{2}}\left(  x\right)  dx
\end{array}
\right]
\end{array}
\right\}
\end{align*}

We want to calculate the probability that $2$ or more deciders make the - choice. There are four quantities to calculate:%
\[
P\left(  -,-,-\right)  ,P\left(  -,-,+\right)  ,P\left(  -,+,-\right)
,P\left(  +,-,-\right)  .
\]

However, note that the last two are the same, due to the symmetry of the $a$
deciders. Hence,
\[
P_{majority\text{ }-}=P\left(  -,-,-\right)  +P\left(  -,-,+\right)
+2P\left(  -,+,-\right).
\]

\bibliographystyle{siamplain}
\bibliography{Caginalp_Doiron}
\end{document}